\documentclass{article}
\usepackage{arxiv}
\usepackage[utf8]{inputenc} 
\usepackage[T1]{fontenc}    
\usepackage{hyperref}       
\usepackage{url}            
\usepackage{booktabs}       
\usepackage{amsfonts}       
\usepackage{nicefrac}       
\usepackage{microtype}      
\usepackage{lipsum}		
\usepackage{graphicx}
\usepackage{doi}
\usepackage{float}
\usepackage{array}
\usepackage{enumitem}
\usepackage{eurosym}
\usepackage[square,sort,comma,numbers]{natbib}

\title{Building a Global Astrotourism Community of Practice Through Astronomy for Development}

\author{ \href{https://orcid.org/0000-0002-9745-0504}{\includegraphics[scale=0.06]{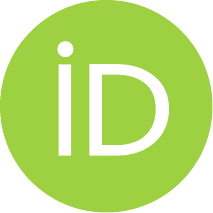}\hspace{1mm}Joyful E. Mdhluli\textsuperscript{1,2}}{ on behalf of the IAU Office of Astronomy for Development} \thanks{Visit our website, https://astrotourism.astro4dev.org/ or email: astrotourism@astro4dev.org} \\
	$^{1}$International Astronomical Union's Office of Astronomy for Development,\\
	$^{2}$South African Astronomical Observatory, Cape Town, 7925, South Africa\\
	\texttt{joy@astro4dev.org} \\
}

\renewcommand{\shorttitle}{\it{Global Astrotourism Community of Practice}}

\hypersetup{
pdftitle={Building a Global Astrotourism Community of Practice Through Astronomy for Development},
pdfsubject={astro-ph.IM},
pdfauthor={Joyful E. Mdhluli},
pdfkeywords={Astrotourism, Dark sky tourism, Astronomy for Development, Community-based tourism, Sustainable Development, Community of Practice},
}

\begin{document}
\maketitle

\begin{abstract}
Astrotourism is an emerging interdisciplinary field situated at the intersection of astronomy, tourism, cultural heritage, and sustainable development. Despite growing global interest, the field remains loosely defined and under-theorized, with limited empirical studies documenting practitioner perspectives and community--led initiatives. This paper presents findings from the first Astrotourism Community Exchange hosted by the International Astronomical Union Office of Astronomy for Development (OAD), which brought together over 200 participants from Africa, Asia, Europe, the Middle East, and Latin America - including practitioners, researchers, educators, tourism professionals, and community developers. Using qualitative thematic analysis of a recorded online community session, we explore how practitioners conceptualize astrotourism, the models being implemented across diverse contexts, and the key challenges shaping the field's development.

Five major themes emerged: (1) astrotourism as a tool for community development, (2) cultural heritage and knowledge systems, (3) environmental sustainability and dark sky protection, (4) accessibility and inclusive design, and (5) conceptual and methodological gaps in the field. Across all themes, astrotourism was consistently framed not merely as a form of niche tourism, but as a hybrid socio-cultural and environmental practice with significant implications for sustainable development and science engagement.
We argue that astrotourism is evolving into a global community of practice rather than a fixed tourism category, and we highlight the need for further interdisciplinary research, data collection frameworks, and inclusive policy development to support its growth.

\end{abstract}


\section{Introduction}       
Astrotourism, broadly defined as tourism centered on experiences of the night sky, astronomy, and related cultural and scientific practices, is rapidly gaining global attention (\cite{fayos1}, \cite{collison}). From dark sky reserves and observatory visits to culturally embedded sky traditions and astrophotography, the field is expanding in both scope and significance. However, despite this growth, astrotourism remains conceptually fragmented. It is variously described as dark sky tourism, night sky tourism,  astronomy tourism, astro-cultural tourism, or observatory-based tourism, with no universally accepted definition or framework (\cite{hobkirk}, \cite{rodrigues}). As one researcher participating in the Community Exchange put it plainly: \textbf{\textit{"Currently we don't even have a real conceptual understanding of what astrotourism is"}}. This definitional ambiguity reflects the genuinely interdisciplinary nature of the field, which spans astronomy, tourism studies, cultural heritage, environmental conservation, and science communication.\\

\begin{figure}[H]
    \centering
    \includegraphics[width=0.2\textwidth]{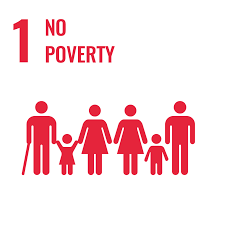}
    \includegraphics[width=0.2\textwidth]{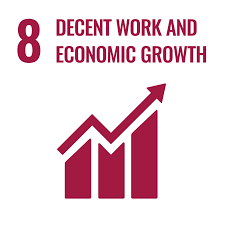}
    \includegraphics[width=0.2\textwidth]{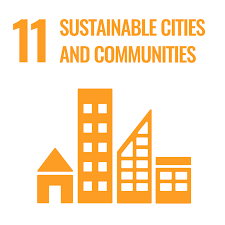}
    \includegraphics[width=0.2\textwidth]{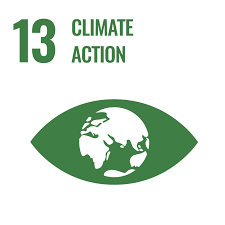}
    \caption{Sustainable Development Goals (SDGs) aligned with astrotourism. Image Credit: \href{https://sdgs.un.org/goals}{United Nations Sustainable Development Goals}}\label{sdgs}
\end{figure}

Early studies positioned astrotourism primarily as a niche or special-interest tourism activity (\cite{collison}), defined by visitors' specific motivation to observe celestial phenomena. Fayos-Sol\`{a} and colleagues (2014) offered a more integrative definition, describing astrotourism as leisure activities that draw on night landscapes as a natural resource while contributing to knowledge sharing, human capital formation, and local development. More recent literature has pushed this further, recognizing astrotourism as a vehicle for sustainable rural development (\cite{rodrigues}, \cite{laeticia}), regenerative tourism, and conservation incentivization (\cite{DarkSky}). At the same time, astrotourism is increasingly being positioned as a tool for sustainable development, particularly in rural and marginalized regions. It offers opportunities for local economic development, cultural preservation, environmental awareness, and educational engagement - all priorities aligned with the United Nations \href{https://sdgs.un.org/goals}{Sustainable Development Goals} (SDGs) (\cite{sdgs}), particularly SDG 1 (No Poverty), SDG 8 (Decent Work and Economic Growth), SDG 11 (Sustainable Cities and Communities), and SDG 13 (Climate Action), see Figure \ref{sdgs}.\\

These alignments are increasingly finding formal policy expression. South Africa's National Astro-Tourism Strategy and Implementation Plan (2023-2033) - jointly developed by the Department of Tourism and the Department of Science, Technology and Innovation, launched in September 2024, and the first such national strategy on the African continent - articulates three strategic pillars: indigenous celestial narratives and human capacity development; infrastructure development; and inclusive tourism growth and partnerships (\cite{dsti}). Its launch in Carnarvon, in the heart of the Northern Cape's dark sky country, and its subsequent presentation to Parliament in February 2025, reflect a broader institutional shift in which astrotourism is transitioning from a practitioner-driven niche into a government-endorsed development priority. That a national government has formalised this transition - embedding it within a ten-year implementation plan spanning science, tourism, and heritage - signals that the field's practitioner-level evolution documented in this paper is already finding its way into policy.

Within this context, the International Astronomical Union Office of Astronomy for Development (IAU OAD) has supported a range of astrotourism initiatives globally since 2013 through its annual call for proposals and, more recently, through a dedicated Astrotourism Flagship Project. To date, the OAD has funded astrotourism-related projects in 17 countries and invested over \euro 100,000 in this focus area, as part of a broader portfolio of 254 projects across 113 countries (\cite{comment2}, \cite{IAU OAD}).\\

Despite this growing ecosystem, there is limited literature documenting how practitioners themselves conceptualize astrotourism and how knowledge is shared across regions and disciplines. This paper addresses that gap by presenting qualitative findings from the first Global Astrotourism Community Exchange, an event designed explicitly to facilitate practitioner-to-practitioner knowledge exchange at the global scale. The aims of this study are: 
\begin{itemize}
    \item to document the emerging perspectives and practices of astrotourism practitioners across diverse geographic and cultural contexts;
    \item to identify key themes and tensions shaping the field's development;
    \item to explore the potential of astrotourism as a vehicle for sustainable development and science engagement.
\end{itemize}

\subsection{Positionality}
The author served as host of the Community Exchange and as coordinator of the OAD Flagship Projects, and specifically the OAD Astrotourism project. This insider position provided privileged contextual knowledge that enriched interpretation, but also introduces a potential confirmation bias toward programmatic framing. Reflexivity was maintained throughout analysis, and findings are presented to reflect the full diversity of participant perspectives, including critiques and tensions expressed during the session.

Additionally, the IAU OAD is represented on the National Astro-Tourism Governance Body responsible for implementing South Africa's National Astro-Tourism Strategy. This institutional relationship is acknowledged where relevant in the analysis.

\section{Background: The IAU OAD Astrotourism Flagship}       
Astronomy serves as a powerful tool for addressing global challenges and promoting sustainable development in various domains, including scientific, technological, social, cultural, economic, and environmental aspects. At first glance, astronomy may seem disconnected from the challenges we face on Earth - poverty, inequality, education, or climate change. But astronomy is far more than the study of stars and galaxies. It inspires curiosity, creativity, and hope. It encourages young people to dream bigger, strengthens education, connects cultures, and reminds us that we all share the same sky. In a world often divided, astronomy gives us perspective - showing us not only how small we are in the universe, but also how connected we are to one another.

Recognizing the profound cultural and historical significance of astronomy, OAD utilizes its versatility to address diverse societal challenges. The OAD was established in 2011 as a joint initiative of the International Astronomical Union (IAU) and the South African National Research Foundation (NRF) with support from the Department of Science, Technology and Innovation (DSTI). Its mandate is to leverage astronomy - including its practitioners, infrastructure, and knowledge - as a tool for development (\cite{IAU}). This is primarily implemented through funding and coordinating projects that use astronomy as a tool to address issues related to sustainable development. In addition to this, the OAD also supports three flagship projects across the thematic areas of socioeconomic development through astronomy, science diplomacy, and astronomy skills for development.\\

The Astrotourism Flagship Project, which falls under the socioeconomic development pillar (\cite{socioeconomic}), is structured around four key programmatic pillars: 
\begin{itemize}
    \item economic development through astronomy-inspired tourism; 
    \item cultural preservation and social inclusion;
    \item environmental sustainability and dark sky protection;
    \item bridging inequality divides through shared access to the night sky.
\end{itemize}

The Flagship project supports community-driven astro-experiences that create local economic opportunities, celebrate Indigenous and cultural sky knowledge, and promote protection of dark skies and natural environments. At the same time, astrotourism can be seen as a way to connect people across social and economic divides through a shared sense of humanity under the stars.

The project is operationalized through an ecosystem model comprising resources, training, implementation support, and community (\cite{ecpsystem}). The astrotourism ecosystem includes open-access resources (\cite{resources}), a publicly available \href{https://astrotourism.astro4dev.org/online-course/}{online training course}, tailored in-person capacity-building workshops, and an annual funding mechanism through which project proposals are competitively selected (the OAD Annual Call for Proposals). A growing online community of practice (\cite{cop}), hosted on the \href{https://astrotourism.astro4dev.org/community/}{Discord} platform, connects practitioners globally across language and geography.\\

This institutional infrastructure provided the foundation for the Community Exchange, which was conceived as a mechanism to strengthen the horizontal knowledge-sharing dimensions of the ecosystem, connecting practitioners with each other, not only with the OAD.

\section{Methods}
\subsection{Study Design}
This study adopts a qualitative exploratory design based on a global online community-of-practice exchange. The event was convened by the OAD as an open virtual engagement aimed at connecting practitioners, researchers, and stakeholders working in astrotourism and related fields.
The purpose of the exchange was to facilitate knowledge-sharing rather than to test a hypothesis, making it suitable for inductive thematic analysis (\cite{inductive}, \cite{inductive1}).

\subsection{Data Collection}
Data were collected from a recorded live online session hosted via Zoom and publicly streamed on YouTube. The dataset comprises both qualitative and limited quantitative components, enabling a mixed-source exploratory analysis of participant discourse and engagement.

The primary dataset includes:
\begin{itemize}
    \item Full video recording\footnote{Youtube Recording available \href{https://www.youtube.com/live/-sSpT9xeO1I?si=Hzw8jpa6NmW7wPcB}{here}}. of the community exchange (1.5 hours)
    \item Spoken presentations from invited practitioners
    \item Moderated discussion segments
    \item Open participant contributions
    \item Organizer observations and notes
\end{itemize}
In addition to the qualitative recording, Zoom polling data was used as a supplementary quantitative input during the session. These polls were designed to capture real-time participant perspectives on key topics related to astrotourism practice and development.

The Zoom poll responses provided:
\begin{itemize}
    \item rapid feedback on thematic priorities (e.g., sustainability, cultural heritage, accessibility, economic development),
    \item indicators of participant interest in different astrotourism applications,
    \item and a broad sense of perceived challenges and opportunities in the field.
\end{itemize}

While not statistically representative due to self-selection and voluntary participation, the poll data served as a contextual triangulation tool that complemented the thematic analysis by providing additional insight into participant priorities, perceived challenges, and practitioner experiences within the astrotourism sector.

Over 200 participants registered for the event, representing a global audience across Africa, Asia, Europe, the Middle East, and the Americas. The combination of recorded dialogue, chat interaction, and Zoom polling enabled a multi-layered dataset capturing both structured and emergent perspectives within the astrotourism community.

\subsection{Data Processing}
The recording was reviewed in full and segmented into thematic sections:
\begin{itemize}
    \item Opening remarks and framing
    \item Case study presentations
    \item Open discussion
    \item Closing reflections
\end{itemize}
A transcript (automated and manually corrected where needed) was used for analysis.

\subsection{Limitations}
This study is based on a single community exchange event and therefore represents a snapshot of a rapidly evolving field. Participation was self-selected and reflects individuals already engaged in astronomy-related practice, those with language barriers to English, limited digital access, or no prior connection to the OAD ecosystem may be underrepresented. The absence of particular organizations, regions, or issues should not be interpreted as their lack of relevance to the broader astrotourism landscape. Findings are not statistically generalizable but are intended as exploratory empirical documentation of an emerging global practice.

\subsection{Ethical Considerations and Data Handling}
This study is based on a publicly streamed and recorded \href{https://astrotourism.astro4dev.org/astrotourism-community-exchange-2026/}{online community exchange} (\cite{community}). Although the session was open to all participants, individual identifying information has been handled in accordance with standard qualitative research ethics practices.

\paragraph{Naming Convention for Contributions:} This paper distinguishes between invited speakers and general participants in accordance with their mode of participation and consent context. Invited speakers who delivered formal presentations during the Community Exchange are identified by name, as their contributions were pre-arranged, intentionally public-facing, and form part of the documented program of the event.

In contrast, contributions from general participants during open discussion sessions, chat interactions, and polling activities have been anonymized to ensure confidentiality and protect the privacy of individuals participating in an open community forum. This includes removal of all personally identifiable information from quotes and thematic analysis.

This distinction reflects standard practice in qualitative research involving public events, where formal presenters are cited as contributors while audience members remain de-identified unless explicit consent for attribution has been obtained.

The full session recording is publicly available via the official livestream link, ensuring transparency and reproducibility of the analysis. However, individual registration data collected for event participation is not publicly shared, in line with standard data protection and privacy considerations for online event registration systems.

Data used in this study are therefore derived from publicly available recorded material and anonymized thematic analysis of spoken and chat contributions.

\section{Findings}
\subsection{Global Participation and Reach}
\begin{figure}[H]
    \centering
    \includegraphics[width=0.45\textwidth]{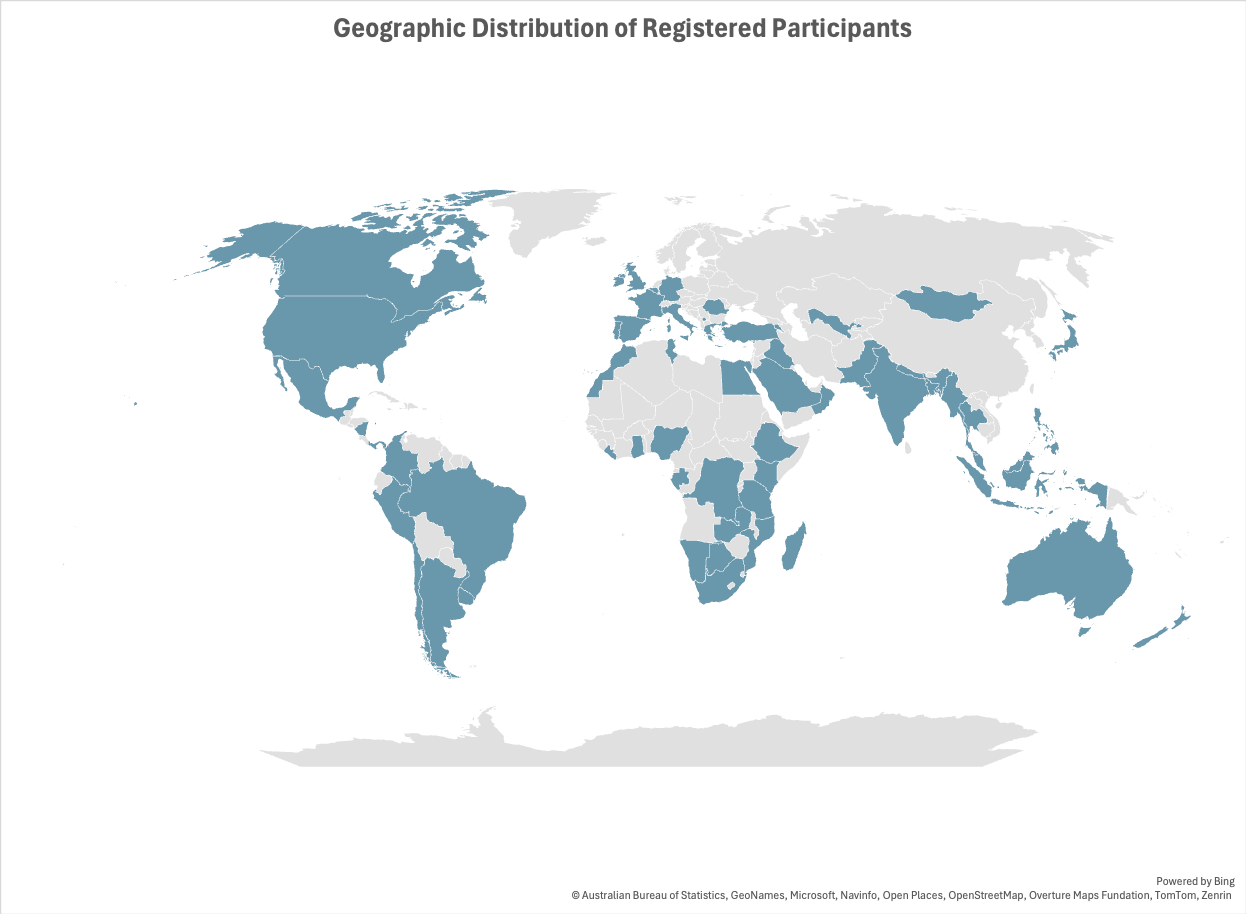}
    \includegraphics[width=0.45\textwidth]{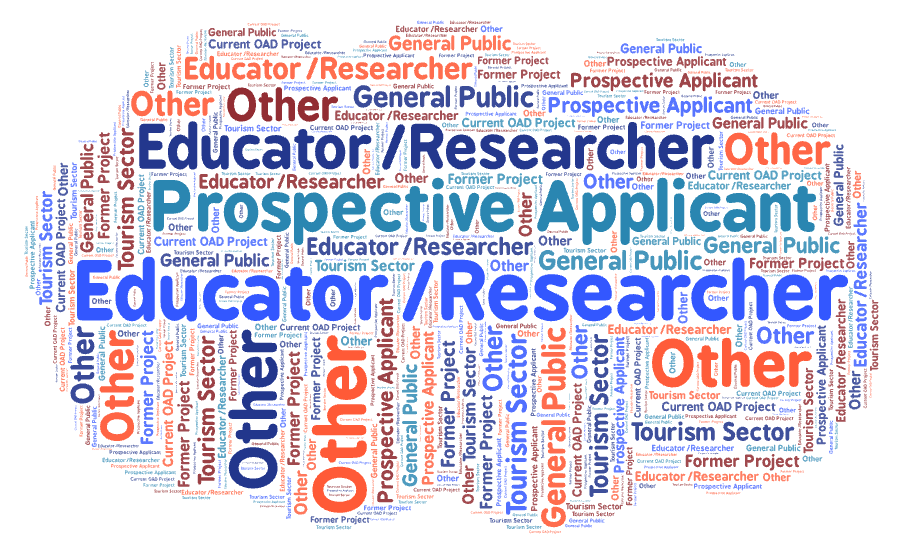}
    \caption{Left: Geographic distribution of registered participants. Right: Identification of participants according to sectors. This information was extracted from the Zoom poll and registration information}\label{participation}
\end{figure}

The Community Exchange brought together over 200 participants representing 59 countries/regions from Africa, Asia, Europe, the Middle East, and Latin America (see Figure \ref{participation} on the left). In the opening poll, participants identified primarily as practitioners, researchers, educators, and tourism professionals (see Figure \ref{participation} on the right. The geographic diversity confirmed a global appetite for knowledge-sharing in astrotourism - a field whose practitioners have historically operated in relative isolation from one another.

As the Community Exchange was open to all interested participants, the OAD had limited control over the disciplinary, professional, and regional representation within the session. To better understand potential gaps within the emerging astrotourism community of practice, participants were asked the following reflective question during the session:\\

\begin{quote}\centering
\textit{"Who is missing in this (virtual) room?"}
\end{quote}

The purpose of this question was to identify stakeholders, sectors, and voices that participants believed should be more actively included in future astrotourism discussions, collaborations, and events. The responses highlighted a strong desire for broader interdisciplinary and institutional engagement within the field.

Participants identified the need for greater participation from:
\begin{itemize}
    \item tourism policymakers and government stakeholders,
    \item representatives from the tourism industry,
    \item environmental and dark sky protection authorities,
    \item business leaders and development investors,
    \item local communities and community elders,
    \item cultural practitioners and arts experts,
    \item industrial partners,
    \item funding bodies, 
    \item and youth from rural communities with access to dark skies.
\end{itemize}

Several participants also emphasized the importance of involving legislators and policymakers in regions heavily affected by light pollution, noting the role such stakeholders could play in strengthening environmental awareness and policy development.

The responses suggest that participants view astrotourism as inherently interdisciplinary and recognize the importance of collaboration across science, tourism, culture, governance, environmental management, and community development sectors. These findings also indicate a growing awareness that the long-term sustainability of astrotourism initiatives depends on inclusive stakeholder participation beyond the astronomy community alone.

\subsection{Practitioner Presentations: Three Case Studies}
One of the highlights of the exchange was hearing directly from projects and practitioners working on the ground in different regions. Three invited presentations offered detailed accounts of OAD-funded astrotourism projects at different stages of implementation. These case studies serve as empirical anchors for the thematic analysis that follows.

\subsubsection{AstroTribe, India (Shweta Kulkarni, Maharashtra)}
Shweta Kulkarni, a science communicator with over 15 years of experience in astronomy outreach based in Pune, Maharashtra, presented the \href{https://astro4dev.org/overview-guide-training-workshop-for-tribal-students/}{AstroTribe} project, funded by the OAD in 2021 and 2022. Her account offered perhaps the most developed conceptual reframing of astrotourism in the session. Kulkarni described a fundamental shift in her understanding of the field through her work in rural India: \textit{"I used to think that astrotourism was basically about helping people look up... But while working in rural India I slowly realized that something unexpected was happening. The people who already lived under the dark skies - the extraordinary skies - were often unable to look up at all because they were busy looking for jobs, water, roads, basic amenities."}\\

This observation led her to reframe the core question of astrotourism from "how do we bring astronomy to communities?" to "can astrotourism help communities build a future underneath the stars?". Through the AstroTribe model, local youth were trained as astroguides, traditional storytelling was integrated into astronomy programming, and science communication became a vehicle for skill development and economic opportunity. The impact extended beyond the project itself. Kulkarni reported that the ecosystem approach \textit{"helped initiate conversations at the state level around dark sky preservation and astrotourism development"} in Maharashtra, culminating in the development of 11 pilot dark sky sites for potential international certification, and public acknowledgment by the state's chief minister of dark sky preservation as a priority.\\

Her most-cited insight from the session distilled a decade of experience into a single principle: \textit{"People do not protect dark skies because astronomers ask them to. They protect them when astrotourism creates ownership."}

\subsubsection{Starlight Bridges, Greece (Dr Margarita Metaxa, Central Greece)}
Dr Margarita Metaxa presented 2026 IAU OAD funded \href{https://astro4dev.org/starlight-bridges-community-astrotourism-through-the-artemis-space-observatory/}{Starlight Bridges} project in Karpenisi, Central Greece - a mountainous, rurally underdeveloped region facing population aging, school closures, the decline of winter tourism under climate change, gender inequality in the workforce, and weak STEM education (Science, Technology, Engineering and Mathematics - STEM). The project leverages the region's exceptional dark skies as an anchor for community and economic development through the Artemis Space Observatory.\\

Activities completed at the time of the exchange included two astrotraining workshops targeting local youth, women, and educators; translation and localization of the OAD Astrotourism Toolkit \cite{resources}; and restoration work at the Artemis Observatory. Upcoming activities included a Dark Sky Festival in August 2026 and STEM outreach in schools from September to November. A particularly innovative component was the design of a "starlight path" through Karpenisi featuring an augmented reality solar system walk - a model that blends astronomical science with experiential tourism.\\

Metaxa raised three questions that generated significant discussion: "How can the projects be sustainable?", "Can astrotourism maintain authenticity if a lot of tourists are coming?" and "How can a dark sky initiative strengthen community cohesion and environmental awareness when it is combined with astrotourism?". These questions framed the tension between scale and integrity that recurred throughout the community discussion.

\subsubsection{AstroDev SA, South Africa (Monique Garden)}
Monique Garden presented the 2026 IAU OAD funded \href{https://astro4dev.org/astro-tourism-for-development-storytelling-science-and-socio-economic-growth-in-south-africa/}{AstroDev SA} project, which targets the Northern Cape and Karoo regions of South Africa - areas home to some of the darkest skies in the world and globally significant astronomy infrastructure including the Southern African Large Telescope (SALT) and the MeerKAT. Despite these world-class assets, Garden noted that \textit{"astrotourism is still largely seen as an add-on experience rather than a fully integrated tourism driver."}

This gap is precisely what South Africa's National Astro-Tourism Strategy and Implementation Plan (2023–2033) is designed to address (\cite{dsti}). AstroDev SA's work in the Northern Cape is therefore situated within a broader national policy framework that formally recognizes astrotourism as a tool for economic development, indigenous knowledge preservation, and inclusive rural tourism growth.

The project, currently at the beginning of its implementation phase, aims to transform astronomy from something visitors are passively aware of into something they actively experience through "storytelling, science communication, local culture and community-led experiences." Phase 1 focuses on mapping astrotourism opportunities in the Northern Cape, developing a digital resource hub, and building a central database of astroguides and tourism stakeholders.

Garden summarized the project's philosophical foundation in terms that echoed both Kulkarni's and Metaxa's: \textit{"It can't just be about astronomy. It's about people. It's about places and it's about the possibilities that we can create for these people."}

\subsection{Emergent Themes}
Five major themes emerged from the community exchange that have consistently been a topic of discussion in other gatherings that the author has been a part of. These themes are listed below:
\paragraph{Astrotourism as a Tool for Community Development:} Participants consistently described astrotourism as a mechanism for local economic empowerment. Case studies demonstrated models involving: training of local youth as astro-guides, community-based tourism enterprises, rural destination development, and integration into local tourism economies. Astrotourism was framed as a driver of inclusive economic opportunity rather than a purely observational activity.

\paragraph{Cultural Heritage and Knowledge Systems:} Astrotourism was strongly linked to cultural identity and heritage. Participants highlighted: Indigenous and traditional sky knowledge systems, storytelling and cultural interpretation, and integration of arts and creative practice. However, concerns were raised about cultural commodification and the need for ethical, community-led engagement.

\paragraph{Environmental Sustainability and Dark Sky Protection:} Dark sky preservation emerged as a central concern. Participants noted that astrotourism can: raise awareness of light pollution, support conservation initiatives, and encourage sustainable infrastructure planning.  A reciprocal relationship was identified between tourism development and environmental protection.

\paragraph{Accessibility and Inclusion:} Accessibility was identified as an underdeveloped area within astrotourism practice. Key issues included: inclusive design for visually impaired audiences, neuro-diversity considerations, physical access to sites, and linguistic and socio-economic barriers. Participants emphasized the need for intentional design frameworks to ensure inclusivity.

\paragraph{Conceptual and Methodological Gaps:} A lack of conceptual clarity was repeatedly highlighted. Key gaps included: absence of a shared definition of astrotourism, limited empirical data on impact, lack of standardized evaluation metrics, and insufficient interdisciplinary integration. This was identified as a barrier to field development.

\subsection{Participant Priorities and Perceived Sector Needs}
During the session, participants were also invited to respond to a series of open-ended Zoom poll questions designed to capture perceptions of challenges, partnerships, and capacity needs within the astrotourism sector. These questions explored:
\begin{itemize}
    \item perceived barriers to astrotourism development,
    \item valuable institutional and community partnerships, and
    \item skills and resources required to strengthen the field.
\end{itemize}

\paragraph{Challenges Facing Astrotourism:}Participants identified several recurring challenges affecting astrotourism development across different regions. The most frequently cited concerns included light pollution, limited institutional support, insufficient funding, geographic remoteness, and low public awareness of astronomy-related tourism opportunities.

Several participants also highlighted challenges associated with infrastructure limitations, safety concerns, and difficulties in sustaining long-term community participation.

\paragraph{Valuable Partnerships and Collaborations:}Participants emphasized the importance of cross-sector partnerships in supporting astrotourism initiatives. Frequently mentioned collaborators included local municipalities, conservation organizations, universities, travel agencies, astronomy associations, and government tourism departments.

The responses suggest that astrotourism development depends heavily on multi-stakeholder collaboration spanning tourism, science, environmental management, and community development sectors.

\paragraph{Skills and Capacity Needs:}Participants identified a wide range of skills and capacity gaps within the astrotourism sector. Frequently cited needs included astronomy communication, storytelling, public engagement, telescope operation, entrepreneurship training, and foundational astronomy education.

Soft skills, particularly communication and audience engagement, were repeatedly emphasized alongside technical astronomy competencies. Several participants also highlighted the importance of empathy and inclusive facilitation practices when working with diverse audiences.

\section{Discussion}
The findings of this study suggest that astrotourism is undergoing rapid conceptual expansion. While traditionally associated with stargazing experiences or dark sky tourism, the field is increasingly being redefined by practitioners as a multi-dimensional development platform.

\subsection{From Tourism Product to Development Ecosystem:}
A key insight is the shift from astrotourism as a niche tourism product toward a broader socio-economic and cultural ecosystem. In multiple contexts, astrotourism is being used to support rural development, create employment opportunities, and strengthen local identity.
This aligns with broader trends in regenerative and community-based tourism, but extends them through the integration of astronomical knowledge and science engagement.

Kulkarni articulated the developmental logic most explicitly, noting that in rural development contexts, wonder alone is insufficient: \textit{"In developing countries, wondering alone is not enough. Governments are balancing hunger, migration, infrastructure, energy, employment. When I spoke about the stars, the room admired the idea. But when we spoke about tourism, local jobs, skill development, regional identity - the room actually began listening differently."}

This pattern of using astronomy as entry point, and economic relevance as the lever for change, was echoed across multiple contexts. A participant from Thailand's National Astronomical Research Institute (NARIT) described 17 years of sustained public awareness-building before astrotourism became viable: \textit{"Before us there was nothing - astronomy happened in Thailand not at all whatsoever. So if we were to do astrotourism right there that would not make any sense because nobody even thought about looking at the stars."} NARIT now serves over one million people annually through five regional observatories, demonstrating how long-term institutional investment in awareness can create the demand conditions for astrotourism to thrive.

Another participant, operating an astrotourism business in Kenya, described a similar logic of integration across conservation, science, and community economy, noting that the interconnectedness of these domains \textit{"is very important to me."} These accounts align with the broader community-based tourism literature, which identifies local ownership, employment creation, and participatory decision-making as the preconditions for sustainable outcomes (\cite{goodwin}; \cite{mitchell}). What is distinctive in the astrotourism cases is the specific role of the night sky as a common resource; accessible from rural and remote areas, free of geographic competition, and uniquely powerful as a shared human heritage.

The findings suggest that practitioners are actively reimagining astrotourism beyond its origins as a niche or special-interest tourism product. In all three featured case studies, and across many contributions from the open discussion, astrotourism was framed not as a service to deliver to tourists but as an ecosystem to build with and for communities. This is a meaningful departure from early scholarly framings, which emphasized the tourist's motivation and the astronomical resource (\cite{fayos1}; \cite{collison}). What practitioners are describing is closer to what the community-based tourism literature calls participatory destination development - where communities are not objects of tourism but agents of it (\cite{mitchell}). The night sky becomes both the draw and the commons around which new forms of local agency are organized.

Kulkarni's formulation - \textit{"instead of bringing tourism to villages, we began helping villages become part of tourism itself"} - encapsulates this shift with particular clarity. Its significance lies in what it changes about the relationship between science, conservation, and development: when communities have economic ownership of the dark sky as a resource, conservation ceases to be an externally imposed constraint and becomes a locally rational choice.
 
\subsection{The Role of Cultural Context}
The study highlights that astrotourism is deeply shaped by local cultural frameworks. Rather than being a standardized model, it adapts to Indigenous knowledge systems, storytelling traditions, and local environmental contexts. Multiple participants described the integration of local storytelling, cosmologies, and practices into their programming as central to creating meaningful and distinctive experiences.  However, this also raises ethical questions around representation, ownership, and commodification of cultural knowledge. 

One participant, presenting from a context combining astronomy outreach with circus arts (juggling and clowning) and currently working with the Din\'{e} people in Arizona, described the power of bringing indigenous constellation knowledge into tourism experiences: \textit{"What we are really enhancing with this is not astronomy but empathy... as soon as you explain the meaning a bug has about the creation of the universe, everyone there is going to buy their arts and crafts. And with this, you are doing what the OAD is truly looking for - not only astronomy or astrotourism, but using it for local development."}

Another participant, developing astrotourism on Sabu Island in Indonesia, presented the concept of "starfishing" - an experience that emulates ancient fishermen's use of stellar navigation, teaching tourists about celestial wayfinding while connecting them to maritime heritage.

However, alongside these examples of meaningful integration, concerns about cultural commodification were clearly articulated. A question was posed during the discussion that framed a key ethical tension in the field: \textit{"At what stage does the cultural knowledge get commodified?"}  How do rural practitioners maintain authentic experiences while also selling them?

An operator of the Cusco Planetarium in Peru - a 19-year family business now listed among National Geographic's 100 best nights of a lifetime -  offered a clear warning: \textit{"We don't have to fall into cultural appropriation. We have to be respectful with our ancestors. We have to respect the cultural heritage... we have to do a lot of research, and in the name of marketing"}. Emphasis was placed that scientists should be seen as allies by tourism practitioners rather than obstacles, and that growing together requires honest engagement with cultural boundaries.
This tension between cultural integration and commodification is well-documented in the indigenous tourism literature (\cite{ryan}), where commercialization of knowledge can erode the spiritual and communal meaning of practices. What the Community Exchange illustrated is that astrotourism practitioners are actively grappling with these questions in real time, without yet having established disciplinary norms or ethical frameworks to guide them.

The Community Exchange demonstrated that cultural heritage is not a decorative addition to astrotourism - it is, in many contexts, the primary substance of the experience. Indigenous navigation traditions (example from Indonesia), mythological storytelling (example from Greece), and community cosmologies (example with Din\'{e}/Navajo context) all emerged as sources of meaning that distinguish authentic astrotourism from commodified stargazing.

At the same time, the commodification concern reveals a fundamental tension that the field must address. When cultural knowledge becomes a tourism product, it enters circuits of value exchange that can erode its original meaning, spiritual function, and community ownership. What is needed in astrotourism is the development of ethical frameworks, perhaps community-negotiated protocols for knowledge sharing, that allow cultural integration without appropriation. The Community Exchange did not resolve this tension, but it did name it clearly, which is a necessary precondition for resolution.

\subsection{Environmental Co-Benefits}
Astrotourism has potential dual benefits: economic development and environmental conservation. The emphasis on dark sky protection illustrates how tourism can incentivize environmental stewardship, particularly in rural regions. However, this relationship requires careful management to avoid unintended environmental impacts from tourism growth

One of the most analytically interesting findings is Kulkarni's account of dark sky conservation becoming politically viable, entering state-level policy discussions, only after it had been connected to economic development arguments. This is not a unique phenomenon; the conservation literature has long recognized that environmental protection is most durable when it aligns with local economic interest (\cite{honey}). What the astrotourism case adds is a specific mechanism: the night sky, because it is most abundant precisely in the places that are most economically marginalized, creates a natural alignment between conservation and development interests that other environmental resources do not always offer.

The Alqueva Dark Sky Reserve in Portugal illustrates what this alignment can produce at scale: the world's first Starlight Tourist Destination certification, recognition as a World Responsible Tourism Award winner in 2021 and 2022, and a model of integrated rural development that has attracted academic attention globally (\cite{rodrigues}).

South Africa's National Astro-Tourism Strategy (\cite{dsti}) gives institutional weight to precisely this dynamic. Its three pillars -  indigenous knowledge and capacity development, infrastructure, and inclusive tourism growth - map directly onto the conservation-development alignment that Kulkarni described in the Indian context, suggesting that practitioners and policymakers are independently converging on the same logic: that protecting dark skies becomes viable when communities have an economic stake in doing so. That this convergence has now produced a formal ten-year government policy framework reinforces the paper's broader argument that astrotourism is not only evolving at the practitioner level, but beginning to reshape how governments think about the intersection of science, heritage, and rural development.

The cases documented in the Community Exchange - Maharashtra, Karpenisi, the Karoo - suggest that similar trajectories are emerging across very different cultural and institutional contexts.
 
\subsection{Inclusion as an Emerging Priority}
Accessibility remains a developing area within astrotourism practice. The findings indicate growing awareness of the need to design inclusive experiences, but limited implementation frameworks currently exist. The conversation initiated by one of the participants, represents a genuine blind spot in the current astrotourism literature and practice. Disability inclusion is well-established as a requirement in tourism design generally (\cite{unwto}), and sensory substitution techniques, such as sonification, have been demonstrated in informal astronomy outreach contexts. But systematic integration of these approaches into astrotourism practice, training, and resources has not occurred. This is inconsistent with the field's explicit commitment to social inclusion and bridging inequality, and represents a clear priority for future development.
 
\subsection{Towards a Global Community of Practice}
Perhaps the most significant finding is the emergence of a global astrotourism community of practice. The exchange itself functioned as a platform for knowledge sharing, collaboration, and conceptual development across regions.

The call for a stronger disciplinary and conceptual foundation echoes across the literature and maps directly onto gaps the practitioners themselves identified. Based on the Community Exchange findings, there is room for a research agenda for the field focusing on:
\begin{itemize}
    \item Definitional consensus: Interdisciplinary scholarly dialogue toward an agreed taxonomy that can accommodate the diversity of astrotourism practice - from observatory visits to ethnological sky traditions.
    \item Impact evaluation frameworks: Standardized instruments for measuring the social, economic, and environmental outcomes of astrotourism initiatives, adaptable to different resource and contextual settings.
    \item Data infrastructure: Regional and global mechanisms for collecting, storing, and sharing participation and impact data.
    \item Cultural ethics: Community-developed frameworks for ethically integrating indigenous and traditional knowledge into tourism programming.
    \item Accessibility and inclusive design: Applied research into multi-sensory and universally designed astrotourism experiences.
\end{itemize}

This suggests that astrotourism is not only a field of practice but also a developing networked ecosystem.

\subsection{Participant Poll Responses}
The Zoom poll responses - based on the participants present in the meeting and reflecting the opinions of a few from around the world - further reinforced the interdisciplinary nature of astrotourism, with participants consistently identifying policymakers, conservation authorities, tourism stakeholders, and local communities as essential collaborators in the development of sustainable astrotourism ecosystems. The questions asked in the Zoom poll and the responses are summarized in Table \ref{tab:pollthemes}.

\begin{table}[h]
\centering
\caption{Summary of Zoom Poll Questions from the Astrotourism Community Exchange}
\label{tab:pollthemes}
\begin{tabular}{p{5cm} | p{9cm}}
\hline
\textbf{Poll Question} & \textbf{Main Response Themes} \\
\hline
What is the biggest challenge for astrotourism? &
Light pollution, limited government support, funding constraints, remoteness of dark sky locations, safety concerns, low public awareness, lack of equipment and infrastructure. \\
\hline
What partnerships have been most valuable? &
Local municipalities, tourism departments, astronomy associations, universities, conservation organizations, travel agencies, local governments, and community stakeholders. \\
\hline
What skills/resources are most needed? &
Astronomy communication, storytelling, telescope operation, public engagement, entrepreneurship training, instrumentation skills, accessibility awareness, and foundational astronomy education. \\
\hline
\end{tabular}
\end{table}

\section{Conclusion}
This paper has presented insights from the first global Astrotourism Community Exchange hosted by the IAU Office of Astronomy for Development on 27 May 2026. Through qualitative thematic analysis of the full session transcript, five themes were identified: astrotourism as a tool for community development, cultural heritage and knowledge systems, environmental sustainability and dark sky protection, accessibility and inclusion, and conceptual and methodological gaps. The overarching finding is that astrotourism is undergoing rapid and deliberate conceptualization at the practitioner level. Across geographically and culturally diverse contexts - from rural Maharashtra to Central Greece, from the Karoo to Sabu Island - practitioners are independently arriving at a similar conclusion: that the field's value is not primarily about delivering astronomical experiences to tourists, but about building systems through which communities can use the night sky as a platform for development, cultural expression, and environmental stewardship.

The Zoom poll responses further reinforced the interdisciplinary and community-centred nature of astrotourism. Participants consistently identified the need for stronger collaboration across tourism, government, conservation, education, and community sectors, while also highlighting significant capacity gaps in communication, storytelling, accessibility, and technical astronomy skills. These findings suggest that the future development of astrotourism will depend not only on dark skies and astronomical infrastructure, but also on inclusive partnerships, local engagement, and sustained investment in human capacity development.

Kulkarni's observation that \textit{"I no longer see astrotourism simply as tourism built around astronomy - I see it as a bridge between science and society, between rural communities and opportunity, and between preservation and development"} may serve as an appropriate conceptual anchor for the field as it matures. If this practitioner-led conceptual evolution is to be supported, it requires the scholarly infrastructure that was identified: interdisciplinary collaboration, shared definitions, empirical data, and ethical frameworks. The Community Exchange, and the network it catalyzed, represents a beginning. As Samyukta Manikumar offered in closing: \textit{"This is just the beginning of a discussion, not the end of it."}

\section{Acknowledgements}
The author thanks all participants of the 2026 Astrotourism Community Exchange for their contributions, and in particular Shweta Kulkarni, Dr Margarita Metaxa, and Monique Garden for volunteering their time and insights as featured presenters. The OAD team and Samyukta Manikumar who provided essential logistical and intellectual support for the event. 

The Office of Astronomy for Development (OAD) a joint initiative of the International Astronomical Union (IAU) and the South African National Research Foundation (NRF) with support from the Department of Science, Technology and Innovation (DSTI).

\end{document}